\documentclass[]{jfm}

\usepackage{upmath}
\usepackage{empheq}
\usepackage{graphicx}
\usepackage{newtxtext}
\usepackage{newtxmath}
\usepackage{natbib}
\usepackage{hyperref}
\hypersetup{
    colorlinks = true,
    urlcolor   = blue,
    citecolor  = black,
}

\newcommand{\RomanNumeralCaps}[1]

\newcommand{\pt}{\partial_t}
\newcommand{\px}{\partial_x}
\newcommand{\py}{\partial_y}
\newcommand{\ps}{\partial_s}
\newcommand{\ihat}{\mathbf{\hat{e}}_x}
\newcommand{\jhat}{\mathbf{\hat{e}}_y}
\newcommand{\nhat}{\mathbf{\hat{n}}}
\newcommand{\ue}{\mathrm{e}}
\newcommand{\ui}{\mathrm{i}}
\newcommand{\ud}{\mathrm{d}}
\newcommand{\bs}{\boldsymbol{\eta}}
\newcommand{\A}{A}

\newcommand{\pxs}{(\px \eta_0)}
\newcommand{\ppxs}{(\px^2 \eta_0)}
\newcommand{\pppxs}{(\px^3 \eta_0)}

\newcommand{\sx}{\eta_x}
\newcommand{\sy}{\eta_y}
\newcommand{\sox}{\eta_{0 x}}
\newcommand{\soy}{\eta_{0 y}}
\newcommand{\sxp}{\eta_x'}
\newcommand{\syp}{\eta_y'}
\newcommand{\soxp}{\eta_{0 x}'}
\newcommand{\soyp}{\eta_{0 y}'}
\newcommand{\sxpp}{\eta_x''}
\newcommand{\sypp}{\eta_y''}
\newcommand{\soxpp}{\eta_{0 x}''}
\newcommand{\soypp}{\eta_{0 y}''}

\newcommand{\bsp}{\bs'}
\newcommand{\bspp}{\bs''}

\title{Modeling Surface Wave Propagation over Meniscus and Scattering by a Surface-piercing Barrier}

 \author{Guoqin Liu, Zhengwu Wang,
 \and Likun Zhang
   \corresp{\email{zhang@olemiss.edu}}}

\affiliation{National Center for Physical Acoustics and Department of Physics and Astronomy, University of Mississippi, University, MS 38677, USA}

\begin{document}
\maketitle

\begin{abstract}
Recent experiments have revealed that the meniscus formed near a surface-piercing barrier can significantly alter the propagation and scattering of capillary-gravity surface waves, beyond what classic flat-surface models predict. 
In particular, wave transmission increases as the barrier pulls up the meniscus, then drops sharply if the barrier is raised further, overturning the meniscus. 
Motivated by these findings, this paper develops two linearized theoretical frameworks that incorporate meniscus geometry and pinned contact line conditions into capillary-gravity wave propagation and scattering. 
Model~1 assumes a vertical wave perturbation of the unperturbed meniscus, thereby extending classic flat-surface boundary conditions in a relatively straightforward manner. 
Model~2 takes a more comprehensive approach, defining surface wave perturbations normal to the curved meniscus and reparameterizing boundary conditions in terms of arc length. 
While Model~1 proves convenient mathematically, it is restricted to single-valued meniscus shapes. 
By contrast, Model~2 is capable of describing multi-valued or overturning free surfaces, thereby capturing a wider range of physically realistic scenarios. 
Numerical simulations based on both models reproduce the experimental observations on how wave transmission varies with changes in meniscus height and contact angles as the barrier is lifted, underscoring the critical influence of meniscus curvature in small-scale wave–structure interactions. 
These results establish a robust theoretical foundation for predicting and optimizing capillary-gravity wave scattering in microfluidic, industrial, and scientific applications.
\end{abstract}


{\bf MSC Codes }  {\it(Optional)} Please enter your MSC Codes here

\section{Introduction}
\label{sec:Introduction}

Fluid surface waves are typically considered on a flat liquid interface, except for capillary oscillations on droplets \citep{vukasinovic2007dynamics,ref:Fayzrakhmanova-2009,sharp2012resonant}, bubbles \citep{harazi2019acoustics}, and liquid bridges \citep{ref:Marr-Lyon2001,morse1996capillary}. These capillary geometries are relatively simple, either spherical or cylindrical in shape. However, when a surface-piercing structure is present, a meniscus naturally forms around the solid boundary, curving the free surface of the liquid near the three-phase (liquid, solid, and air) contact lines and deforming it into an interface with a spatially varying curvature under the influence of gravity. Although the significance of meniscus effects in small-scale fluid systems has been explored, their influence on surface wave scattering has not been investigated until recent experimental studies \citep{Guillaume2016, zhengwu2025}. Such meniscus-involved scattering relates to phenomena commonly observed in daily life, such as when wind-generated ripples encounter surface-piercing obstacles, and it can play a role in small-scale fluid control and containment systems. However, theoretical analyses of capillary-gravity wave scattering have thus far been limited to a flat free surface. The present work develops a theoretical framework that explicitly incorporates the meniscus geometry into the propagation and scattering of capillary-gravity waves.

Capillary-gravity waves arise where both gravity and surface tension significantly influence fluid motion, bridging two classical regimes of wave behavior—gravity waves on larger scales driven by gravity and capillary waves on smaller scales driven by surface tension. 
Although the scattering of gravity waves by barriers has been extensively studied in contexts such as coastal engineering, fluid containment, and wave energy systems \citep{Dean1945,Ursell1947,mei1989applied}, capillary-scale scattering typically emerges in smaller fluid configurations where even minimal contact with solid boundaries can alter wave dynamics. 
Examples include laboratory setups and systems designed to stabilize or guide fluid interfaces: liquid globes on wire loops \citep{ref:Pettit-2005}, helical-wire-stabilized liquid cylinders \citep{ref:Lowry-Thiessen-2007}, microstructured surfaces with pillars \citep{ref:Rothstein-2010}, and open capillary channels in biotechnology \citep{berthier2019open}. 

Meniscus effects are pervasive in both normal- and micro-gravity applications and have been studied in a variety of settings.
The geometry of the meniscus not only modifies static fluid configurations \citep{zhang2013,ref:Marr-Lyon2001,ref:Lowry-Thiessen-2007,bostwick2009capillary,bostwick2015stability,berhanu2020capillary}, but also strongly influences wave dynamics, such as the onset of capillary waves \citep{shao2021role}, the frequency shift of traveling waves \citep{ref:Miles-1992,kidambi2004effects,nicolas2005effects,kidambi2009meniscus}, and the stability of large-amplitude or standing waves \citep{perlin2000capillary,nguyem2011effect}. 
For instance, \citet{shao2021role} demonstrated that a curved interface—whether positively or negatively curved—is essential for initiating capillary waves, whereas a perfectly flat interface precludes their formation. 
\citet{kidambi2009meniscus} and \citet{nicolas2005effects} showed that pinned or free menisci can produce more pronounced shifts in wave frequencies and damping than viscous effects. 
Recent work has also explored advanced theoretical formulations for capillary–gravity wave scattering that incorporate dynamic contact line models and associated dissipation mechanisms \citep{guoqin2025}, underscoring how contact line conditions can significantly affect scattering coefficients and phase relationships. 
Altogether, these findings emphasize the importance of accounting for meniscus and contact line phenomena in many practical scenarios involving surface-piercing structures.

Recently, \citet{Guillaume2016} and \citet{zhengwu2025} demonstrated that the meniscus formed near a vertical plate barrier significantly alters the transmitted wave energy—far beyond what flat-surface models predict. Of special note, \citet{zhengwu2025} revealed an intriguing observation: 
when raising the hydrophobically coated surface-piercing barrier such that the meniscus is elevated from a flat surface, the wave transmission even increases. Moreover, As the barrier is lifted further and the meniscus inclines enough to overturn, the transmission drastically decreases, diverging from classical scattering theories that do not incorporate meniscus curvature effects. This discrepancy underscores a critical gap: existing theoretical and numerical models are unable to reproduce these experimental findings without properly accounting for the meniscus geometry and dynamic contact line conditions. To address this shortcoming, an improved framework is needed—one that integrates both the capillary-induced meniscus shape and appropriate boundary conditions at the contact line for a more accurate prediction of capillary-gravity wave propagation and scattering.

\begin{figure}
    \centering
    \includegraphics[width=0.8\textwidth]{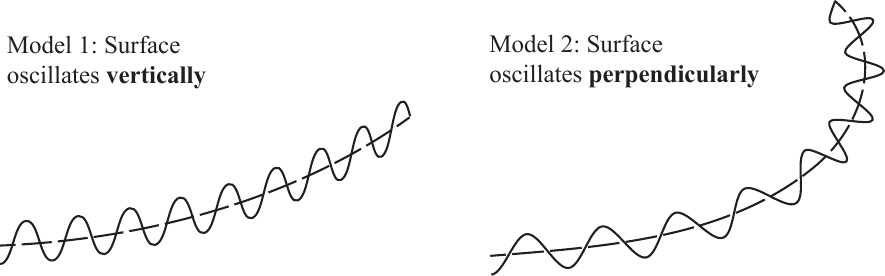}
    \caption{Model comparison.}
    \label{fig:m_comp}
\end{figure}

We propose two linearized models (See Fig.~\ref{fig:m_comp}): Model 1, which extends the classic flat-surface approach by assuming vertical wave displacement along a single-valued meniscus, and Model 2, which accounts for wave perturbations perpendicular to the curved surface and thus handles more complex or overturned meniscus shapes. We then apply these models to study how a vertical barrier interacts with the meniscus, focusing on the experimentally observed peak transmission when the barrier is at the meniscus top and the drastic decline as the barrier is raised further. While Model 1 successfully predicts the initial increase in transmission, it cannot capture the drop once the meniscus becomes multi-valued or overturned. By contrast, Model 2 encompasses the meniscus curvature more robustly and reproduces both the peak and the subsequent sharp decrease, demonstrating that incorporating the meniscus shape is essential for understanding and predicting wave scattering in such small-scale systems.

The remainder of this paper is organized as follows. \S~\ref{sec:M1}, we derive Model 1, formulating the wave-induced displacement as strictly vertical and applying linearized boundary conditions to a single-valued meniscus shape. \S~\ref{sec:M2} presents Model 2, which accommodates displacements perpendicular to the curved surface and thus generalizes to more complex or overturned meniscus geometries. In \S~\ref{sec:Bridge}, we establish the connection between these two models, showing that Model 2 reduces to Model 1 under single-valued meniscus profiles, while Model 1 in turn reverts to classic flat-surface conditions when the unperturbed meniscus height is zero. \S~\ref{sec:Scattering} details the simulation setup for capillary-gravity wave scattering by a surface-piercing barrier and compares the numerical predictions with recent experimental observations. Finally, \S~\ref{sec:Summary} summarizes the main findings and concludes with implications for future research.

\section{Model 1: Vertical Oscillations}
\label{sec:M1}

Consider the two-dimensional motion of an inviscid, incompressible, and irrotational fluid of density $\rho$. The fluid is subject to gravity $g$ and surface tension $\sigma$. In the fluid domain, we introduce the velocity potential $\phi$ with $\mathbf{v} = \bnabla\phi$, satisfying Laplace’s equation:
\begin{equation}
    \nabla^2 \phi = 0. \label{eq:Laplace_vert}
\end{equation}

Let the free surface be described by $y = \eta(x, t)$. On this free surface, the kinematic boundary condition states that a fluid particle on the surface must remain on it. In differential form, this can be written as:
\begin{equation}
    \pt \eta + \px\phi \px\eta = \py\phi,
\end{equation}
or more compactly in vector form as:
\begin{equation}
    \boxed{
    \nhat \bcdot \bnabla\phi = \frac{\pt \eta}{\sqrt{1+(\px\eta)^2}}, \label{eq:kinematic_gen_M1}
    }
\end{equation}
where the unit normal $\nhat$ pointing out of the fluid is given by: 
\begin{equation}
    \nhat = \frac{-\px \eta \ihat + \jhat}{\sqrt{1+(\px\eta)^2}}. \label{eq:n_M1}
\end{equation}
Physically, \eqref{eq:kinematic_gen_M1} states that the normal fluid velocity at the free surface matches the time rate of change of the interface elevation (including horizontal advection).

Bernoulli’s equation and the pressure balance at the free surface give:
\begin{align}
    \pt\phi + g\eta + \tfrac{1}{2} (\bnabla\phi)^2 + \rho^{-1} p &= 0, \label{eq:dynamic_1} \\
    p &= -\sigma \kappa(x, t), \label{eq:dynamic_2}
\end{align}
where \eqref{eq:dynamic_2} shows that the discontinuity between the fluid pressure $p$ at the surface and the atmospheric pressure  (set to zero for convenience) equals the capillary pressure, which is proportional to the surface tension $\sigma$ and the local curvature of the free surface $\kappa$:
\begin{equation}
    \kappa = -\bnabla \cdot \nhat = \frac{\px^2\eta}{( 1 + (\px\eta)^2 )^{3/2}}.
\end{equation}

Substituting \eqref{eq:dynamic_2} into \eqref{eq:dynamic_1} to eliminate pressure $p$ leads to the dynamic boundary condition: 
\begin{equation}
    \boxed{
    \pt\phi + \tfrac{1}{2} (\bnabla\phi)^2 = \frac{\sigma}{\rho} \frac{\px^2\eta}{( 1 + (\px\eta)^2 )^{3/2}} - g\eta. \label{eq:dynamic_gen_M1}
    }
\end{equation}

Equations \eqref{eq:kinematic_gen_M1} and \eqref{eq:dynamic_gen_M1} form the main governing conditions at the free surface for this problem.

\subsection{Linearized Equations for Capillary-Gravity Waves on a Curved Surface}
\label{ssec:lin_M1}

We now derive a linearized model for small-amplitude waves on a curved free surface under both gravity and surface tension. Let the free surface be:
\begin{subequations}\label{eq:pert}
\begin{equation}
    \eta = \eta_0(x) + \epsilon \eta_1(x, t), 
\end{equation}
where $\eta_0(x)$ is the static (equilibrium) meniscus profile, $\eta_1(x, t)$ is the \emph{vertical} displacement by the waves, and $\epsilon$ is a small dimensionless parameter measuring the relative amplitude of the perturbation. In our model, we assume $|\epsilon| \ll 1$. The choice of vertical displacement will be contrasted later with a formulation in which the displacement is measured normal to $\eta_0$.

Similarly, let the velocity potential be expanded as:
\begin{equation}
    \phi = \epsilon \phi_1(x, y, t).
\end{equation}
\end{subequations}

Substitute \eqref{eq:pert} into the exact kinematic boundary condition \eqref{eq:kinematic_gen_M1}:
\begin{equation}
    \nhat \bcdot \bnabla (\epsilon\phi_1) = \frac{\pt (\eta_0 + \epsilon\eta_1)}{\sqrt{1+(\px(\eta_0 + \epsilon\eta_1))^2}}, \label{eq:kinematic_app3}
\end{equation}
where $\nhat$, as in \eqref{eq:n_M1}, is evaluated at $y = \eta_0 + \epsilon\eta_1$. Since $\eta_0$ is independent of $t$ and $\epsilon\eta_1$ is small, we linearize about $y = \eta_0$. Retaining only first-order terms $\mathcal{O}(\epsilon)$, \eqref{eq:kinematic_app3} becomes:
\begin{equation}
    \boxed{
    \nhat_0 \bcdot \bnabla \phi_1 = \frac{\pt \eta_1}{\sqrt{1+\pxs^2}},
    }
    \label{eq:kinematic_lin_M1}
\end{equation}
where $\nhat_0$ is the normal vector at the static (equilibrium) meniscus profile $y = \eta_0$. Physically, \eqref{eq:kinematic_lin_M1} states that the fluid’s normal velocity at the interface matches the time rate of change of the small vertical displacement $\epsilon\eta_1$.

Next, substitute \eqref{eq:pert} into the dynamic condition \eqref{eq:dynamic_gen_M1}:
\begin{equation}
    \epsilon\pt \phi_1 + \tfrac{1}{2} (\epsilon\bnabla\phi_1)^2 = \frac{\sigma}{\rho} \frac{\px^2(\eta_0 + \epsilon\eta_1)}{( 1 + (\px(\eta_0 + \epsilon\eta_1))^2 )^{3/2}} - g (\eta_0 + \epsilon\eta_1). \label{eq:dynamic_app3}
\end{equation}
We consider the zeroth-order (equilibrium) balance $\mathcal{O}(1)$ and then the first-order terms $\mathcal{O}(\epsilon)$:

At \emph{zeroth order} $\mathcal{O}(1)$, we obtain:
\begin{equation}
    \kappa_0 = \frac{\px^2\eta_0}{( 1 + \pxs^2 )^{3/2}} = \frac{\eta_0}{a^2}, \label{eq:dynamic_app4}
\end{equation}
where $a = \sqrt{\sigma/\rho g}$ is the capillary length. This relation governs the equilibrium meniscus shape $\eta_0(x)$.

At \emph{first order} $\mathcal{O}(\epsilon)$, \eqref{eq:dynamic_app3} becomes:
\begin{equation}
    \boxed{
    \frac{\rho}{\sigma} \pt \phi_1 = \frac{\px^2\eta_1}{( 1 + \pxs^2 )^{3/2}} - \frac{3 \eta_0}{a^2} \frac{\pxs}{ 1 + \pxs^2 } \px\eta_1 - a^{-2} \eta_1.
    }
    \label{eq:dynamic_lin_M1}
\end{equation}

\eqref{eq:kinematic_lin_M1} and \eqref{eq:dynamic_lin_M1} constitute the linearized kinematic and dynamic boundary conditions we derived for small-amplitude capillary-gravity waves, explicitly assuming the perturbation $\eta_1(x, t)$ is a \emph{vertical} displacement from the static meniscus profile $\eta_0(x)$. 

If the unperturbed surface is flat ($\eta_0 = 0$), \eqref{eq:kinematic_lin_M1} and \eqref{eq:dynamic_lin_M1} reduce to
\begin{subequations}\label{eq:SBC-textbook}
\begin{align}
    \py \phi_1 &= \pt \eta_1, \\
    \rho \pt \phi_1 &= \sigma \px^2\eta_1 - \rho g \eta_1,
\end{align}    
\end{subequations}
which are the classical linearized boundary conditions for capillary-gravity waves on a flat surface \citep{fetter2003theoretical}.

\subsection{Implementation to Meniscus Surface}

\begin{figure}
    \centering
    \includegraphics[width=\textwidth]{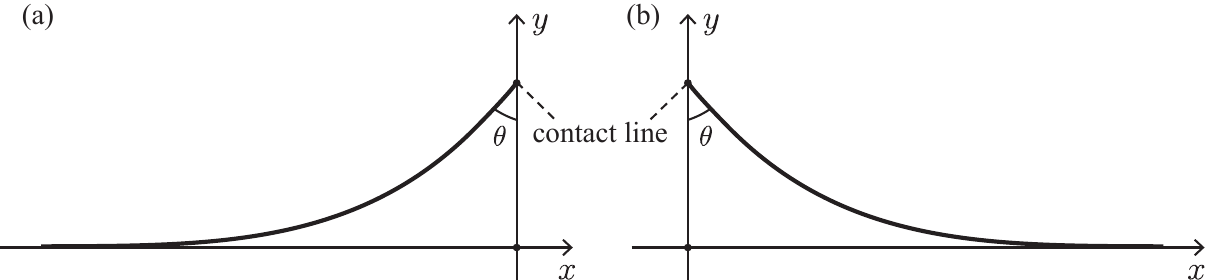}
    \caption{Schematic for two kinds of meniscus: (a) a left-going meniscus, extending from the contact line to \(x \to -\infty\); (b) a right-going meniscus, extending from the contact line to \(x \to +\infty\).}
    \label{fig:meniscus}
\end{figure}
We now implement this formulation to a static meniscus described by $y = \eta_0(x)$, where $\eta_0$ satisfies the static equilibrium balance between gravity and surface tension, as in \eqref{eq:dynamic_app4}.
Following the derivation in \citep{C-W-book, landau2013fluid}, integrating \eqref{eq:dynamic_app4} once yields:
\begin{equation}
    1 - \frac{ \eta^2_0 }{ 2 a^2 } = \frac{1}{ \sqrt{ 1 + ( \px \eta_0 )^2 } }, \label{eq:surf_relation_1.5_M1}
\end{equation}
which can be reorganized as:
\begin{equation}
    \px \eta_0 = \pm \left( - \frac{\eta_0}{a} \frac{\sqrt{1 - \eta_0^2/4a^2}}{1 - \eta_0^2/2a^2} \right), \label{eq:surf_relation_2_M1}
\end{equation}
where the “$\pm$” distinguishes between the two kinds of meniscus:
“$-$” for the \emph{left-going} meniscus (Fig.~\ref{fig:meniscus}(a)), 
“$+$” for the \emph{right-going} meniscus (Fig.~\ref{fig:meniscus}(b)).
Integrating \eqref{eq:surf_relation_2_M1} gives the implicit form for the static meniscus profile $\eta_0(x)$:
\begin{equation}
    a \cosh^{-1} \left( \frac{ 2 a }{ \eta_0 } \right) - \sqrt{4 a^2 - \eta^2_0 } = \pm (x - x_0) , \label{eq:surf}
\end{equation}
\noindent where \( x_0 \) is determined by matching \( \eta_0(x) \) with the capillary rise at the contact lines (denoted by \( h \)), leading to a static contact angle \( \theta \) (measured from the vertical direction to the meniscus),
\begin{equation}\label{eq:theta-h-a}
    \sin\theta = 1 - \frac{ h^2 }{ 2 a^2 },
\end{equation} 
as given by~\eqref{eq:surf_relation_1.5_M1}.

For small-amplitude waves on this meniscus, by substituting \eqref{eq:surf_relation_1.5_M1} and \eqref{eq:surf_relation_2_M1} to eliminate $\px \eta_0$, we reduce the linearized kinematic and dynamic conditions \eqref{eq:kinematic_lin_M1} and \eqref{eq:dynamic_lin_M1} to:
\begin{subequations}\label{eq:bc_men_M1}
    \begin{empheq}[box=\fbox]{align}
        \nhat_0 \bcdot \bnabla \phi_1 &= \left( 1 - \frac{\eta^2_0}{2 a^2} \right) \pt \eta_1 , \label{eq:kinematic_men_M1} \\
        \frac{\rho}{\sigma} \pt \phi_1 &= \left( 1 - \frac{\eta^2_0}{2 a^2} \right) ^3 \px^2 \eta_1 \pm \frac{3 \eta_0^2}{a^3} \left( 1 - \frac{\eta^2_0}{2 a^2} \right) \sqrt{1 - \frac{\eta_0^2}{4a^2}}\; \px \eta_1 - a^{-2} \eta_1 .  \label{eq:dynamic_men_M1}
    \end{empheq}
\end{subequations}
Recall that “$\pm$” distinguishes between the two kinds of meniscus (see Fig.~\ref{fig:meniscus}):
“$+$” for the \emph{right-going} meniscus,  
“$-$” for the \emph{left-going} meniscus. \eqref{eq:bc_men_M1} constitute the kinematic and dynamic boundary conditions we derived for capillary-gravity waves propagating on a meniscus surface under a vertical displacement assumption. The extra terms in \eqref{eq:bc_men_M1}, when comparing with the boundary conditions without meniscus \eqref{eq:SBC-textbook}, are associated with the surface tension term as introduced by the meniscus.

The velocity potential $\phi_1$ itself in \eqref{eq:bc_men_M1} still satisfies Laplace’s equation,
\begin{equation}
    \nabla^2 \phi_1 = 0, \label{eq:laplace_men_M1}
\end{equation}
within the fluid. 

Together, the boundary conditions \eqref{eq:bc_men_M1} for $\eta_1$ we have and the Laplace equation \eqref{eq:laplace_men_M1} for $\phi_1$ form a complete boundary-value problem for small-amplitude capillary-gravity waves on a meniscus, with the wave displacement $\eta_1$ measured in the vertical direction from $\eta_0(x)$.

In reality, free-surface perturbations occur \emph{normal} to the meniscus, particularly where the interface has large slopes or overturning shapes. Requiring the free surface to be expressed as $y = \eta(x, t)$ restricts the geometry to single-valued profiles, which fails if the meniscus has vertical tangents or loops back in the $x$-direction (see Fig.~\ref{fig:m_comp}). These limitations motivate the more general formulation presented in Model 2.

\section{Model 2: Perpendicular Oscillations}
\label{sec:M2}

In the previous section, we expressed each point on the free surface in terms of its horizontal coordinate \(x\) and vertical displacement \(\eta(x,t)\).  This works well so long as the interface can be described by a single-valued function.  However, if the surface becomes steep, overturns, or even loops back on itself, the relationship \(y=\eta(x, t)\) fails to remain single-valued.

Here, we introduce a more general formulation in which each point on the free surface is parameterized by its \emph{arc length}, \(s\), measured from the contact line.  In other words, we switch from the “vertical” description \((x,\eta(x,t))\) to a “curved coordinate” description:
\begin{equation}
    \bs(s, t) = ( \sx (s, t), \sy (s, t) ) = \sx (s, t) \ihat + \sy (s, t) \jhat. \label{eq:eta_M2}
\end{equation}
Here, $s = 0$ corresponds to the contact line on the boundary, and $s$ increases in the direction along the free surface (i.e.\ along the curve). $\sx(s, t)$ and $\sy(s, t)$ are the $x$- and $y$-coordinates of the free surface at arc length $s$ and time $t$.

For simplicity, we denote the derivative with respect to arc length $s$, $\ps$, as a prime ($'$). Since $s$ measures the \emph{length} along the curved interface, the tangent to the surface $\ps\bs$ is automatically a unit vector.  Hence, we have:
\begin{equation}
    \sxp^2 + \syp^2 = 1, \label{eq:sp_relation}
\end{equation}
where $\sxp = \ps \sx$ and $\syp = \ps \sy$.

By switching to \((\eta_x,\eta_y)\) as functions of \(s\), we free ourselves from the restriction that \(\eta\) must be a single-valued function of \(x\).  The interface can now develop overturns, vertical tangents, or even self-intersecting “loops” (though additional constraints typically avoid self-intersections in physical contexts).  This approach naturally accommodates \emph{normal} (perpendicular) displacements from the unperturbed surface, because once we know the tangent direction, we also know the unique normal direction at each point on the interface.  

Figure~\ref{fig:m_comp} illustrates how a parametrized curve can represent both simple and complex surface shapes.  If the surface is nearly flat, it coincides with the usual \(\eta(x)\) approach; if it becomes steep or multi-valued, the arc-length parameterization still applies.

In the rest of this section, we derive the kinematic and dynamic boundary conditions in these arc-length coordinates.  Because the normal and tangent directions are built into the \(\bs(s,t)\) formulation, we can implement the physics of free-surface motion—in particular, capillary–gravity wave behavior—without being limited to a single-valued height function in the \((x,y)\)-plane.

\subsection{General Boundary Conditions at the Free Surface}
\label{ssec:Gen_M2}

For the fluid’s free surface to remain a material boundary, the fluid velocity at the surface must match the velocity of the surface itself. This leads to the kinematic boundary condition:
\begin{equation}
    \nhat \bcdot \bnabla \phi = \nhat \bcdot \partial_t \bs, \label{eq:kinematic_gen_temp1_M2}
\end{equation}
where $\nhat = -\syp\ihat + \sxp\jhat$ is the unit normal to the surface, directed outward from the fluid domain. Substituting $\bs(s, t)$ from \eqref{eq:eta_M2} into \eqref{eq:kinematic_gen_temp1_M2} gives:
\begin{equation}
    \nhat \bcdot \bnabla \phi = \nhat \bcdot \pt \bs = 
    \begin{vmatrix} 
    \sxp & \syp \\ 
    \pt \sx & \pt \sy
    \end{vmatrix}.
\end{equation}
Defining the determinant of two vectors $\boldsymbol{A}$ and $\boldsymbol{B}$ by $\det( \boldsymbol{A}, \boldsymbol{B} ) = \begin{vmatrix} A_{x} & A_{y} \\ B_{x} & B_{y} \end{vmatrix}$, we write the kinematic condition succinctly as:
\begin{equation}\label{eq:kinematic_gen_M2}
    \boxed{
    \nhat \bcdot \bnabla \phi = \det(\bsp, \pt \bs).
    }
\end{equation}

Bernoulli’s equation and the pressure balance at the free surface retain their usual forms:
\begin{align}
    \pt\phi + g \sy + \tfrac{1}{2} (\bnabla\phi)^2 + \rho^{-1} p &= 0, \label{eq:dynamic_3} \\
    p &= -\sigma\kappa(s, t), \label{eq:dynamic_4}
\end{align}
where $\kappa(s, t)$ is the curvature of the surface. In terms of $\sx$ and $\sy$, the curvature $\kappa$ is:
\begin{equation}
    \kappa = \frac{\sxp \sypp - \syp \sxpp}{(\sxp^2 + \syp^2)^{3/2}} = \sxp\sypp - \syp\sxpp = \det(\bsp, \bspp),
\end{equation}
where we have used $\sxp^2 + \syp^2 = 1$, as in \eqref{eq:sp_relation}.

Combining \eqref{eq:dynamic_3} and \eqref{eq:dynamic_4} yields the dynamic boundary condition:
\begin{equation}
    \boxed{
    \pt\phi + \tfrac{1}{2} (\bnabla\phi)^2 = \frac{\sigma}{\rho} \det(\bsp, \bspp) - g \sy. \label{eq:dynamic_gen_M2}
    }
\end{equation}

\eqref{eq:kinematic_gen_M2} and \eqref{eq:dynamic_gen_M2} are the full boundary conditions for a free surface whose displacement is measured along the local normal. In \S~\ref{sec:M1}, we treated the wave displacement as purely vertical; here, we allow the surface to move normal to itself. 

Notably, one can show that if the surface $\bs$ can be expressed with a single-valued function $y = \eta(x, t)$, then \eqref{eq:kinematic_gen_M2} and \eqref{eq:dynamic_gen_M2} reduce to the conditions in Model 1, \eqref{eq:kinematic_gen_M1} and \eqref{eq:dynamic_gen_M1}. Specifically, if $y = \eta(x, t)$, then $\pt\sx = 0$, $\pt\sy = \pt \eta(x, t)$, and $\ps = \px/\sqrt{1+(\px\eta)^2}$. Under these conditions, the normal velocity and curvature in Model 2 become identical to those in Model 1:
\begin{align}
    \nhat \bcdot \bnabla \phi &= \det(\bsp, \pt\bs) = \sxp \pt \sy - \syp \pt \sx \equiv \frac{\pt \eta}{\sqrt{1+(\px\eta)^2}}, \\
    \kappa &= \det(\bsp, \bspp) = \sxp \sypp - \syp \sxpp \equiv \frac{\px^2\eta}{( 1 + (\px\eta)^2 )^{3/2}}. 
\end{align}
Thus, Model 2 is a strict generalization of Model 1, remaining valid even if the interface is multi‐valued in $x$.

\subsection{Linearized Equations for Capillary-Gravity Waves on a Curved Surface}

We now linearize the boundary conditions \eqref{eq:kinematic_gen_M2} and \eqref{eq:dynamic_gen_M2} under the assumption of small‐amplitude wave perturbations measured along the normal to an unperturbed (static) meniscus. Let the unperturbed surface be:
\begin{equation}
    \bs_0(s) = \sox(s) \ihat + \soy(s) \jhat,
\end{equation}
with outward-pointing unit normal $\nhat_0 = - \soyp \ihat + \soxp \jhat$. We write the perturbed surface as:
\begin{align}
    \bs(s, t) &= \bs_0(s) + \epsilon\xi_1(s, t) \nhat_0, \\
    &= (\sox - \epsilon\soyp \xi_1) \ihat + (\soy + \epsilon\soxp \xi_1) \jhat, \label{eq:pert_M2}
\end{align} 
where $\xi_1(s, t)$ is the scalar displacement in the normal direction and the dimensionless scale parameter $\epsilon$ satisfies $|\epsilon| \ll 1$.

Note that $\bs$ is parameterized by the \emph{same} arc length $s$ as $\bs_0$; this reparameterization remains valid under a small‐perturbation assumption $\epsilon\xi_1 \ll a$ (see Appendix~\ref{app:arclength}). The property $\sxp^2 + \syp^2 = 1$ still holds for $\bs$ to first order. Meanwhile, the velocity potential is expanded as:
\begin{equation}
    \phi = \epsilon\phi_1(x, y, t).
\end{equation}

Substituting \eqref{eq:pert_M2} into kinematic boundary condition \eqref{eq:kinematic_gen_M2} and retaining only first-order terms $\mathcal{O}(\epsilon)$ gives:
\begin{equation}
    \boxed{
    \nhat_0 \bcdot \bnabla \phi_1 = \pt{\xi_1}.
    }\label{eq:kinematic_lin_M2}
\end{equation}
Thus, the normal fluid velocity at the unperturbed meniscus matches the time rate of change of the normal displacement $\pt{\xi_1}$, as expected.

Similarly, substituting \eqref{eq:pert_M2} into dynamic boundary condition \eqref{eq:dynamic_gen_M2} yields:
\begin{equation}\label{eq:dynamic_app1_M2}
    \epsilon\pt\phi_1 + \tfrac{1}{2} (\epsilon\bnabla\phi_1)^2 = \frac{\sigma}{\rho}
    \begin{vmatrix}
        (\sox - \epsilon\soyp \xi_1)' & (\soy + \epsilon\soxp \xi_1)' \\
        (\sox - \epsilon\soyp \xi_1)'' & (\soy + \epsilon\soxp \xi_1)''
  \end{vmatrix}
  - g (\soy + \epsilon\soxp \xi_1). 
\end{equation}
At \emph{zeroth order} $\mathcal{O}(1)$, the equilibrium meniscus satisfies:
\begin{equation}
    \kappa_0 = \soxp \soypp - \soyp \soxpp = \soy/a^2, \label{eq:dynamic_app2_M2}
\end{equation}
where $a = \sqrt{\sigma/\rho g}$ is again the capillary length. At \emph{first order} $\mathcal{O}(\epsilon)$, \eqref{eq:dynamic_app1_M2} becomes:
\begin{equation}
    \boxed{
    \frac{\rho}{\sigma} \pt\phi_1 = \xi_1'' - \frac{1}{a^2} \left( \soxp + \frac{2 \soy^2}{a^2} \right) \xi_1.
    }\label{eq:dynamic_lin_M2}
\end{equation}

\eqref{eq:kinematic_lin_M2} and \eqref{eq:dynamic_lin_M2} constitute the linearized kinematic and dynamic boundary conditions for small‐amplitude waves whose displacement $\xi_1(s, t)$ is measured \emph{normal} to the unperturbed meniscus $\bs_0(s)$.

\subsection{Implementation to Meniscus Surface}

Finally, we apply this perpendicular‐displacement formulation to a meniscus whose equilibrium shape $\bs_0(s)$ is parameterized by arc length. In Appendix~\ref{app:Meniscus_M2}, explicit expressions for $\sox(s)$, $\soy(s)$ and $\kappa_0(s)$ are derived:
\begin{subequations}\label{eq:bs0_M2}
    \begin{align}
    \sox(s) &= \pm \left\{ s + 4 a \left[ \frac{1}{1 + \A^2} - \frac{1}{1 + \A^2 \ue^{-2 s/a}}\right] \right\},\label{eq:bs0x_M2} \\
    \soy(s) &= \frac{4 a \A \ue^{-s/a}}{1 + \A^2 \ue^{-2 s/a}}, \\
    \kappa_0(s) &= \frac{4 a^{-1} \A \ue^{-s/a}}{1 + \A^2 \ue^{-2 s/a}},
\end{align}
\end{subequations}
where $\A = \tan(\tfrac{\pi}{8} - \tfrac{\theta}{4})$, $\theta$ is the static contact angle, and the sign “$\pm$” here distinguishes between the two kinds of meniscus:
“$-$” for the \emph{left-going} meniscus (Fig.~\ref{fig:meniscus}(a)), 
“$+$” for the \emph{right-going} meniscus (Fig.~\ref{fig:meniscus}(b)).

Substituting \eqref{eq:bs0x_M2} into the linearized kinematic and dynamic boundary conditions \eqref{eq:kinematic_lin_M2} and \eqref{eq:dynamic_lin_M2} to eliminate $\sox$, one obtains:
\begin{subequations}\label{eq:bc_men_M2}
\begin{empheq}[box=\fbox]{align}
    \nhat_0 \bcdot \bnabla \phi_1 &= \pt{\xi_1}, \label{eq:kinematic_men_M2}\\
    \frac{\rho}{\sigma} \pt\phi_1 &= \xi_1'' - \frac{1}{a^2} \left[ \pm \left( 1 - \frac{\soy^2}{2 a^2} \right) + \frac{2 \soy^2}{a^2} \right] \xi_1. \label{eq:dynamic_men_M2}
\end{empheq}
\end{subequations}
Recall that the sign “$\pm$” here distinguishes between the two kinds of meniscus:
“$-$” for the \emph{left-going} meniscus, 
“$+$” for the \emph{right-going} meniscus. \eqref{eq:bc_men_M2} constitute the kinematic and dynamic boundary conditions we derived for capillary-gravity waves propagating on a meniscus surface under a normal displacement assumption. In the fluid domain, $\phi_1$ still satisfies Laplace’s equation,
\begin{equation}
    \nabla^2 \phi_1 = 0. \label{eq:laplace_men_M2}
\end{equation}

\eqref{eq:bc_men_M2} together with the Laplace equation \eqref{eq:laplace_men_M2} in the fluid, complete the boundary‐value problem for small‐amplitude capillary‐gravity waves on a meniscus with normal displacement $\xi_1$.

Compared to the purely vertical displacement in Model 1, measuring the wave amplitude along the normal direction $\nhat_0$ provides a more physically realistic description, especially for steep menisci or geometries with significant curvature. Indeed, if the meniscus has vertical tangents or a multi‐valued shape in $x$, Model 2 remains valid, whereas Model 1 does not.

\section{Bridging the Linearized Boundary Conditions from Model 2 to Model 1}
\label{sec:Bridge}

In this section, we compare the linearized free‐surface boundary conditions in Model 2 and Model 1, denoting the respective surface perturbations by $\xi_1$ (Model 2) and $\eta_1$ (Model 1).  We show that when a certain small parameter (involving the equilibrium profile $\eta_0$) is negligible, Model 2’s linearized conditions reduce to those of Model 1. Importantly, we limit on the case that the unperturbed meniscus $\bs_0 = \sox \ihat + \soy \jhat$ can be represented by a single‐valued function $y = \eta_0(x)$. Otherwise, as discussed in Section~\ref{ssec:Gen_M2}, the general boundary conditions of Model 2 and Model 1 do not coincide, and neither will their linearizations.

\subsection{Kinematic Boundary Condition}

Recall that in Model 2, the linearized kinematic boundary condition \eqref{eq:kinematic_lin_M2} is:
\begin{equation}
  \nhat_0 \bcdot \bnabla \phi_1 = \pt \xi_1, \label{eq:app2_eq1}
\end{equation}
while in Model 1, the corresponding condition \eqref{eq:kinematic_lin_M1} is:
\begin{equation}
  \nhat_0 \bcdot \bnabla \phi_1 = \frac{\pt \eta_1}{\sqrt{1+\pxs^2}}. \label{eq:app2_eq2}
\end{equation}
Here, $\xi_1$ represents a normal displacement (Model 2), while $\eta_1$ represents a vertical displacement (Model 1). Geometrically, one can show:
\begin{equation}
    \xi_1 \ud s = \eta_1 \ud x, \label{eq:app2_eq3}
\end{equation}
which implies:
\begin{equation}
    \xi_1 = \eta_1 \ps \sox = \frac{\eta_1}{\sqrt{1 + \pxs^2}}. \label{eq:app2_eq4}
\end{equation}
Thus, moving the free surface “vertically” by $\eta_1$ at some horizontal coordinate $x$ is locally equivalent to moving it “normally” by $\xi_1$, when the interface slope is $\px \eta_0$. 
From \eqref{eq:app2_eq4}, it follows that:
\begin{equation}
    \pt \xi_1 = \frac{\pt \eta_1}{\sqrt{1+\pxs^2}}, \label{eq:app2_eq5}
\end{equation}
showing that the linearized kinematic condition \eqref{eq:app2_eq1} in Model 2 is indeed identical to \eqref{eq:app2_eq2} in Model 1, once $\xi_1$ and $\eta_1$ are related through the geometry of the surface.

\subsection{Dynamic Boundary Condition}

The difference arises in the dynamic boundary conditions. In Model 2, the linearized dynamic condition \eqref{eq:dynamic_lin_M2} reads:
\begin{equation}
    \frac{\rho}{\sigma} \pt\phi_1 = \ps^2\xi_1 - \frac{1}{a^2} \left( \ps \sox + \frac{2 \soy^2}{a^2} \right) \xi_1, \label{eq:app2_eq6} 
\end{equation}
whereas in Model 1, the corresponding condition \eqref{eq:dynamic_lin_M1} reads:
\begin{equation}
    \frac{\rho}{\sigma} \pt \phi_1 = \frac{\px^2\eta_1}{( 1 + \pxs^2 )^{3/2}} - \frac{3 \eta_0}{a^2} \frac{\pxs}{ 1 + \pxs^2 } \px\eta_1 - a^{-2} \eta_1. \label{eq:app2_eq7}
\end{equation}
Using the relation \eqref{eq:app2_eq4} ($\xi_1 = \eta_1/\sqrt{1 + \pxs^2}$) and $\ps = \px/\sqrt{1+(\px\eta)^2}$, we can transform the second‐derivative term $\ps^2 \xi_1$ in \eqref{eq:app2_eq6} to an expression of $\eta_1$:
\begin{align}\label{eq:app2_eq9}
    \ps^2 \xi_1 &= \frac{1}{\sqrt{1 + \pxs^2}} \px \left( \frac{1}{\sqrt{1 + \pxs^2}} \px \left( \frac{\eta_1}{\sqrt{1 + \pxs^2}} \right) \right), \nonumber \\
    &= \left[1 + \pxs^2\right]^{-5/2} \bigg\{ \left[1 + \pxs^2\right] \px^2 \eta_1 - 3 \pxs\ppxs \px\eta_1 \nonumber \\
    &\qquad\qquad\qquad + \left( 4 \pxs^2 \ppxs^2 \left[1 + \pxs^2\right]^{-1} - \left[ \ppxs^2 + \pxs\pppxs \right] \right) \eta_1 \bigg\}.
\end{align}
Using the equilibrium condition \eqref{eq:dynamic_app4} to eliminate higher-order derivatives of $\eta_0$, \eqref{eq:app2_eq9} becomes:
\begin{equation}
    \ps^2 \xi_1 = \frac{\px^2\eta_1}{( 1 + \pxs^2 )^{3/2}} - \frac{3 \eta_0}{a^2} \frac{\pxs}{ 1 + \pxs^2 } \px\eta_1 - a^{-2} \left( \frac{\eta_0^2}{a^2} \frac{1}{\sqrt{1+\pxs^2}} + \frac{\pxs^2}{1+\pxs^2} \right) \eta_1.
\end{equation}
Consequently, \eqref{eq:app2_eq6} becomes:
\begin{align}
    \frac{\rho}{\sigma} \pt \phi_1 &= \ps^2\xi_1 - \frac{1}{a^2} \left( \frac{1}{\sqrt{1 + \pxs^2}} + \frac{2 \eta_0^2}{a^2} \right) \frac{\eta_1}{\sqrt{1 + \pxs^2}} , \nonumber \\
    &= \frac{\px^2\eta_1}{( 1 + \pxs^2 )^{3/2}} - \frac{3 \eta_0}{a^2} \frac{\pxs}{ 1 + \pxs^2 } \px\eta_1     
    - \underbrace{\frac{3 \eta_0^2}{a^2 \sqrt{1+\pxs^2}}}_{\text{denoted by } \delta_e}  \frac{\eta_1  }{a^2}  
    - \frac{\eta_1  }{a^2},
    \label{eq:app2_eq11}
\end{align}
where on the right hand side, the first three terms are associated with the surface tension, while the last (or fourth) term is associated with the gravity force. This boundary condition \eqref{eq:app2_eq11} differs from \eqref{eq:app2_eq7} by the third term.

Thus, even though the \emph{general} boundary conditions are equivalent (Section~\ref{ssec:Gen_M2}), the \emph{linearized} versions differ by this extra term. This highlights that treating the free‐surface perturbation as purely vertical is mathematically convenient but not strictly rigorous for waves propagating over meniscus where $\eta_0 \neq 0$.

 \subsection{Conditions Under Which Models 1 and 2 Coincide}
 \label{ssec:coincide}

We now ask under what condition these two linearized models produce the \emph{same} results for a single‐valued meniscus profile $y = \eta_0(x)$. Equivalently, we seek when the third term on the right hand side of \eqref{eq:app2_eq12} is negligible. For the convenience of analysis, the factor in front of $\eta_1/a^2$ of this term is now denoted by:
\begin{equation}
    \delta_e = \frac{3 \eta_0^2}{a^2 \sqrt{1+\pxs^2}}. \label{eq:app2_eq12}
\end{equation}
Using \eqref{eq:surf_relation_2_M1}, 
\begin{equation}
    1 - \frac{ \eta^2_0 }{ 2 a^2 } = \frac{1}{ \sqrt{ 1 + ( \px \eta_0 )^2 } },\label{eq:app2_eq14}
\end{equation}
$\delta_e$ in \eqref{eq:app2_eq12} becomes:
\begin{equation}
    \delta_e = 6 \bigg( 1 - \frac{1}{ \sqrt{ 1 + ( \px \eta_0 )^2 } } \bigg) \frac{1}{ \sqrt{ 1 + ( \px \eta_0 )^2 } }. \label{eq:app2_eq15}
\end{equation}
Further, using geometrical relation between contact angle $\theta$ and $\px\eta_0$, $\cot \theta = \px\eta_0$, the extra term $\delta_e$ in \eqref{eq:app2_eq15} becomes a function of contact angle $\theta$:
\begin{equation} \label{eq:delta_e_theta}
    \delta_e = 6 \left( 1 - \lvert\sin\theta\rvert \right) \lvert\sin\theta\rvert. 
\end{equation}
When the meniscus is sufficiently flat ($\left| \px \eta_0 \right| \approx 0$ or $\theta  \approx 90^\circ$), the factor $\delta_e$ approaches zero and the linearized Model 1 and Model 2 coincide.

For extremely large slopes, $\left| \px \eta_0 \right| \approx \infty$, corresponding to contact angles very close to $0^\circ$ or $180^\circ$, the factor $\delta_e$ also approaches zero.  However, one should be cautious about the physical interpretation at these extreme angles. At $0^\circ$ or $180^\circ$, the meniscus surface becomes nearly vertical, making the assumption of vertical perturbation with a pinned contact line physically questionable. 

Overall, when the interface can be expressed as a single‐valued function $y = \eta_0(x)$ with gentle slopes, the difference between Model 1 and Model 2 remains small, because both models use the same linear wave theory. In practice (see Section~\ref{sec:Scattering}), the predictions of the two models often agree well unless the meniscus has substantial curvature or overturning, at which point the purely vertical formulation of Model 1 may fail while Model 2 still applies.

\section{Capillary-Gravity Wave Scattering by a Surface-Piercing Barrier: Experiments and Simulations}
\label{sec:Scattering}

\begin{figure}
    \centering
    \includegraphics[width=0.9\textwidth]{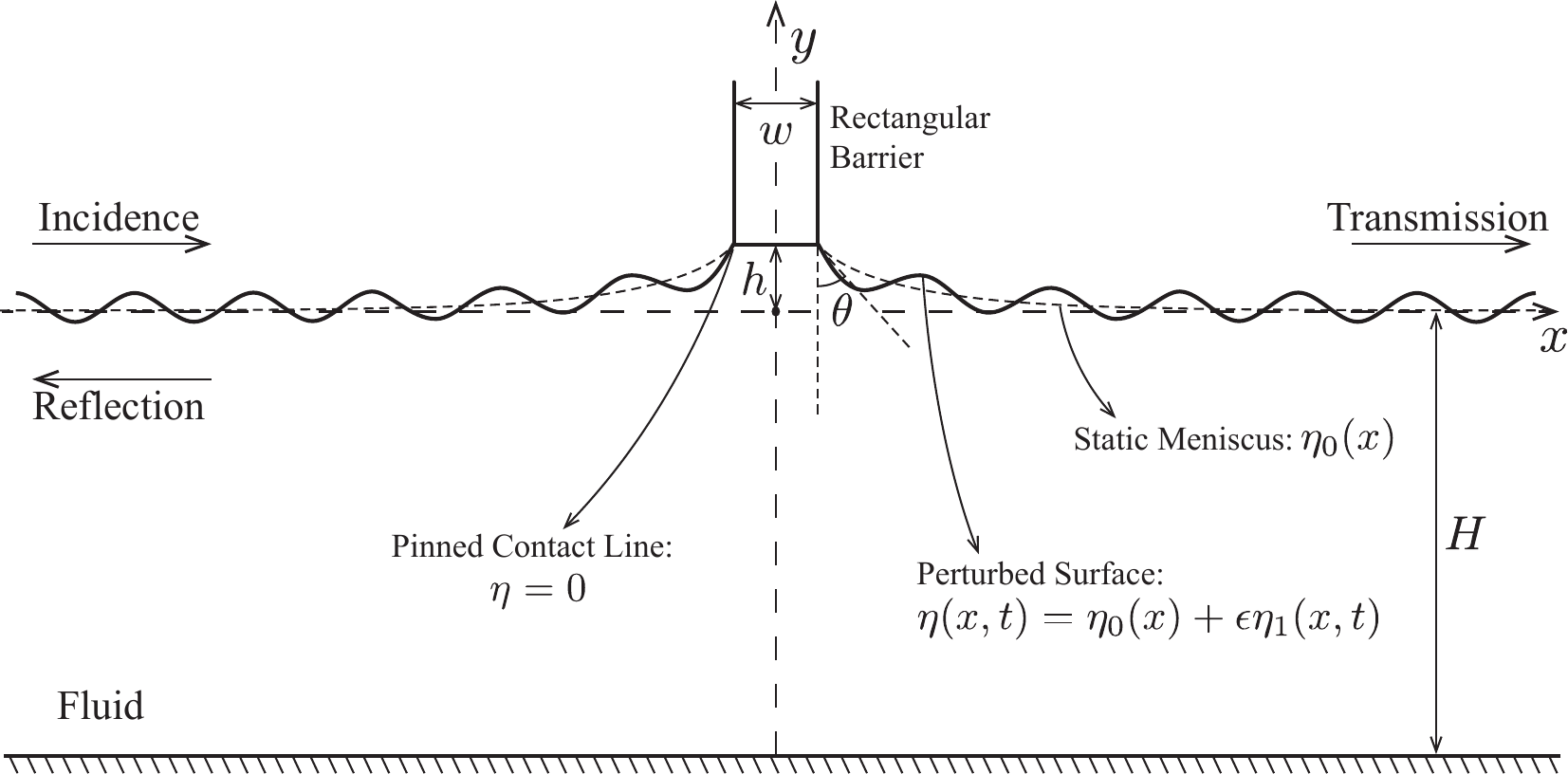}
    \caption{Schematic of a capillary–gravity surface wave scattered by a surface-piercing rectangular plate barrier of thickness $w$.  The barrier’s bottom edge is at a height $h$ relative to the flat fluid surface, and the contact line is pinned along that edge.  The contact angle $\theta$ is measured relative to the barrier’s vertical face.}
    \label{fig:schematic}
\end{figure}

In this section, we apply our two linearized wave models (Model~1 and Model~2) to the problem of capillary–gravity wave scattering by a surface-piercing plate barrier (see Fig.~\ref{fig:schematic}).  This configuration has been studied experimentally by \citet{zhengwu2025}, allowing direct comparison of simulated results against measured wave transmission.

\subsection{Model Formulation for the Scattering Problem}

First, we formulate the complete boundary-value problem for the capillary-gravity wave scattering. We begin by considering an equilibrium configuration in which a vertically placed rectangular barrier of thickness $w$ contacts the fluid of depth $H$ solely along its bottom, located at a height $h$ relative to the fluid surface level away from the meniscus (see Fig.~\ref{fig:schematic}). The contact lines are pinned at the edges of the barrier so the height $h$ is the capillary rise of the meniscus at the contact lines. We will vary $h$ from $-\sqrt{2}a$ to $2a$, corresponding to the contact angle varies from 180$^\circ$ to -90$^\circ$, as follows from the contact angle equation  \eqref{eq:theta-h-a}.

A Cartesian coordinate system $(x, y)$ is introduced, with $x=0$ aligned along the barrier's centerline and $y=0$ coincident with the fluid surface level away from the meniscus. The fluid, characterized by density $\rho$, surface tension $\sigma$, meets the barrier bottom edges at a static contact angle $\theta$, measured relative to the barrier's vertical face. This setup is used to investigate the scattering of time-harmonic capillary-gravity waves—defined by angular frequency $\omega$ and wavenumber $k$—incident from $x \rightarrow -\infty$. The corresponding governing equations and boundary conditions describing the boundary-value problem for this capillary-gravity wave scattering problem are shown as follows:

(a) \emph{Laplace Equation in Fluid Domain.} 
In both Model~1 and Model~2, the velocity potential $\phi_1$ in the fluid domain satisfies Laplace’s equation, as in \eqref{eq:laplace_men_M1} and \eqref{eq:laplace_men_M2}:
\begin{equation}
    \nabla^2 \phi_1 = 0 \quad \text{in the fluid domain.}
\end{equation}

(b) \emph{Free Surface Boundary Conditions.} 
At the free surface $y = \eta_0(x)$, we impose the linearized kinematic and dynamic conditions we derived in Model~1 (Section~\ref{sec:M1}) and Model~2 (Section~\ref{sec:M2}). Specifically, these are given by \eqref{eq:bc_men_M1} for Model~1 and \eqref{eq:bc_men_M2} for Model~2. 

(c) \emph{Pinned Contact Line.}  
At the barrier bottom edges (where the contact lines are pinned), the perturbation $\eta_1$ is set to zero, representing a pinned contact line with no slip:
\begin{equation}
    \eta_1 = 0 \quad \text{at contact lines.} \label{eq:pcl_bc}
\end{equation}

(d) \emph{Solid Boundaries.}  
The barrier bottom at $y=h, x\in [-\tfrac{w}{2}, \tfrac{w}{2}]$, and the fluid bottom at $y=-H$ both enforce a no-penetration condition,
\begin{equation}
    \nhat \bcdot \bnabla \phi_1 = 0,
\end{equation}
where $\nhat$ is the outward normal to the solid boundary.

(e) \emph{Far-Field Form.} 
At the free surface of the fluid of depth $H$, the incident surface wave of angular frequency $\omega$ and wavenumber $k$ has the form of \citep{fetter2003theoretical}:
\begin{subequations}\label{eq:incident}
\begin{align}
\phi_I(x,y,t) 
&= 
\phi_A 
\frac{\cosh(k (y + H))}{\sinh(k H)}
\ue^{\ui(k x - \omega t)}, 
\\
\eta_I(x,t)
&= 
\eta_A \ue^{\,\ui(k x - \omega t)},
\end{align}
\end{subequations}
where amplitude $\phi_A$ and $\eta_A$ have the relation:
\begin{equation}
    \phi_A = -\ui \frac{\omega}{k} \eta_A,
\end{equation}
and $\omega$ and $k$ satisfy the dispersion relation:
\begin{equation}
    \omega^2 = \tfrac{\sigma}{\rho}k^3\,(1 + B)\tanh(kH),
\end{equation}
where
\begin{equation}
    B = \frac{\rho g}{\sigma k^2}, \label{eq:B}
\end{equation}
is the Bond number, characterizing the relative effect between gravity and surface tension.
 
At large distances from the barrier, the surface elevation takes the far-field form:
\begin{subequations}\label{eq:farfield}
\begin{align}
\eta_1 &= \eta_I + \eta_R, &\text{at } x \rightarrow -\infty, \label{eq:farfield_IR}\\
\eta_1 &= \eta_T, &\text{at } x \rightarrow +\infty, \label{eq:farfield_T}
\end{align}
where the reflected and transmitted waves are defined by:
\begin{align}
\eta_R &= R \eta_I (-x, t) = R \eta_A \ue^{ \ui( - k x - \omega t ) } ,\label{eq:eta_R}\\
\eta_T &= T \eta_I (+x, t) = T \eta_A \ue^{ \ui( k x - \omega t ) } .\label{eq:eta_T}
\end{align}
\end{subequations}
Thus, $R$ and $T$ are the complex reflection and transmission coefficients. 

\subsection{Simulation Setup}

We carry out simulations of this scattering problem in a finite domain where we truncate the $x$-direction to $x\in[-L, L]$. Here, $L = 10\lambda$ is a large distance so that we can impose radiation boundary conditions at $x=\pm L$ to simulate the far-field behavior and allow outgoing waves to exit with minimal reflection. At $x = -L$, following from \eqref{eq:farfield_IR}, we impose:
\begin{align}
\partial_x \phi_1 + \ui k \phi_1 &=  2 \ui k \phi_A \frac{\cosh(k(y + H))}{\sinh(kH)} e^{ - \ui k L }, \\
\partial_x \eta_1 + \ui k \eta_1 &=  2 \ui k \eta_A e^{ - \ui k L }. 
\end{align}
At $x = L$, following from \eqref{eq:farfield_T}, we impose:
\begin{align}
\partial_x \phi_1 - \ui k \phi_1 &=  0, \\
\partial_x \eta_1 - \ui k \eta_1 &=  0. 
\end{align}
We then extract $R$ and $T$ by comparing the computed $\eta_1(\pm L)$ with the known incident wave amplitude $\eta_A$:
\begin{align}
    R &= \eta_1(-L) e^{\ui k L} / \eta_A - 1. \\
    T &= \eta_1(L) e^{- \ui k L} / \eta_A.
\end{align}
Due to the pinned contact line condition \eqref{eq:pcl_bc}, the contact line does not slip along the barrier, so there is no frictional energy loss. As a result, we verify numerically that $|T|^2 + |R|^2 = 1$, confirming overall energy conservation. Consequently, we focus on $|T|$ as the primary indicator of how barrier height, contact angle $\theta$, and meniscus geometry influence the wave transmission.

In our simulations, boundary-value problems incorporating both Model 1 and Model 2 are implemented using a finite‐element method in COMSOL Multiphysics\textsuperscript{\textregistered}. The governing equations consist of the Laplace equation for $\phi_1$ in the fluid domain and the dynamic boundary condition for $\eta_1$ at the free surface, coupled through the surface kinematic boundary condition, where the surface perturbation $\eta_1$ acts as the source term for the velocity potential $\phi_1$ field below.

We take water with density $\rho = 997\,\mathrm{kg/m^3}$, surface tension $\sigma = 71.99\,\mathrm{mN/m}$, gravitational acceleration $g=9.8$ m/s$^2$, and fluid depth $H = 9.2\,\mathrm{cm}$.  The barrier thickness is $w=3.18\,\mathrm{mm}$. A wave of frequency $f=15\,\mathrm{Hz}$ implies an angular frequency $\omega = 2\pi f = 94.2\,\mathrm{rad/s}$. The corresponding wavenumber is $k = 408.7\,\mathrm{rad/m}$, giving a Bond number $B \approx 0.813$ and wavelength $\lambda \approx 15.4\,\mathrm{mm}$. The capillary length is $a$= 2.7~mm. These parameters match the experiment by \citet{zhengwu2025}.

A uniform mesh with element size $\Delta = \lambda/25$ in both $x$ and $y$ directions ensures sufficient fine resolution of both the wave field and the near-barrier fluid mechanics. Convergence tests with finer and coarser meshes (doubled and halved mesh densities) indicated that changes in the computed transmission coefficient $|T|$ remained below 1\%, demonstrating sufficient convergence.

\subsection{Simulation Results and Comparisons with Measurements}

\begin{figure}
    \centering
    \includegraphics[width=\textwidth]{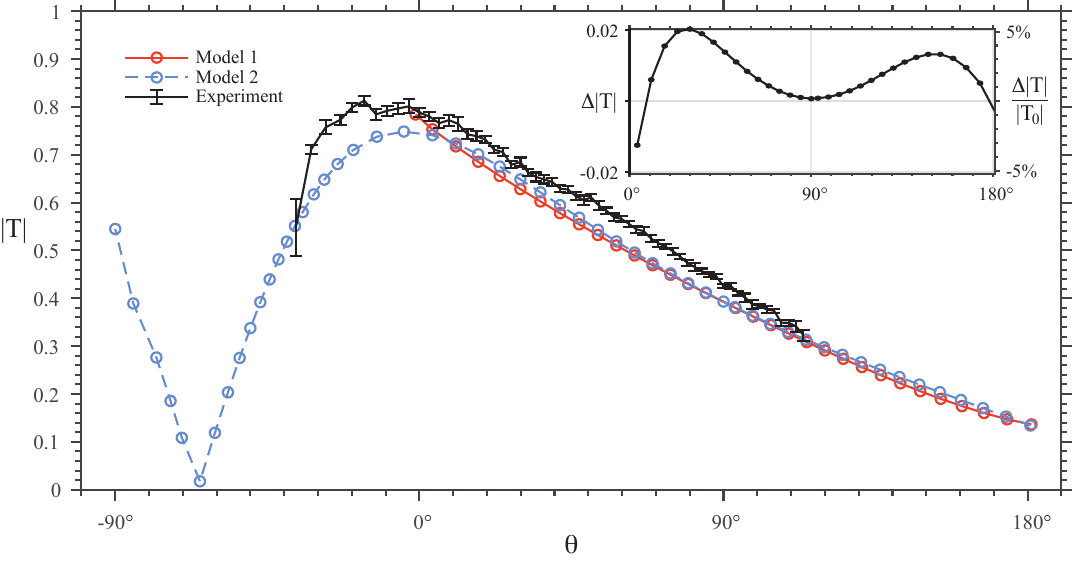}
    \caption{Transmission coefficient $|T|$ versus $\theta$ for a 15 Hz wave interacting with a 3.18~mm-wide, surface-piercing plate barrier. Experimental data \citep{zhengwu2025} are shown by the black solid curve, Model 1 by the red solid line, and Model 2 by the blue dashed line. For $\theta < -36.5^\circ$, the experiment’s meniscus breaks; Model 2 predictions beyond that angle are extrapolated states not observed in the laboratory. The inset shows the difference $\Delta |T| = |T|_\text{Model 2} - |T|_\text{Model 1}$ and the relative difference $\Delta |T| / |T_0|$, where $|T_0|$ is the transmission coefficient from Model 2 at flat meniscus $\theta = 90^\circ$.}
    \label{fig:exp_vs_sim}
\end{figure}

\begin{figure}
    \centering
    \includegraphics[width=\textwidth]{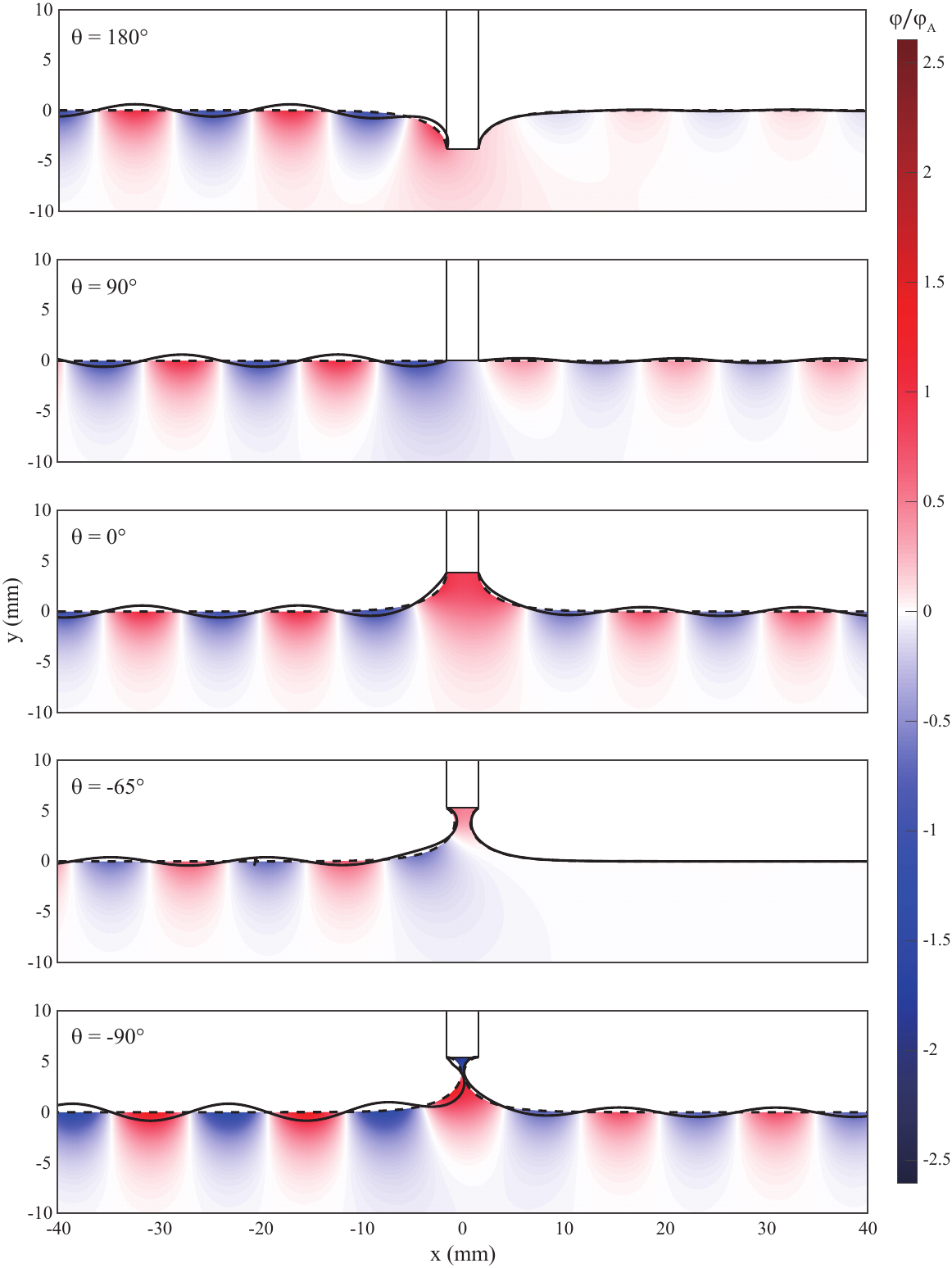}
    \caption{Snapshots of the normalized velocity potential $\phi/\phi_A$ and free-surface profiles (solid black curves) for five characteristic contact angles: $\theta = 180^\circ, 90^\circ, 0^\circ, -65^\circ,$ and $-90^\circ$. The dashed black curves denote the unperturbed meniscus. At $\theta = -65^\circ$, practically all wave energy is reflected, while at $\theta = - 90^\circ$ a slight resurgence of downstream wave amplitude appears.}
    \label{fig:fields}
\end{figure}

In the experiment by \citet{zhengwu2025}, wave transmission $|T|$ is measured as a function of contact angle $\theta$ (see Fig.~\ref{fig:exp_vs_sim}). The barrier is hydrophobically coated and initially positioned so that the pinned contact line forms a static contact angle $\theta = 113.6^\circ$. From this configuration, raising the barrier forces the meniscus to rise along the barrier, reducing $\theta$ below $\theta = 113.6^\circ$. As the meniscus is pulled higher, it may overturn (negative $\theta$) and eventually break once its curvature becomes too large to sustain (at around $\theta \approx -36.5^\circ$). 

Our linearized models do not enforce physical stability limits on $\theta$; instead, we varies continuously from $180^\circ$ to $-90^\circ$. This range encompasses both the experimentally accessible states and additional hypothetical configurations beyond the point of meniscus breakup in the laboratory. Figure~\ref{fig:exp_vs_sim} compares experimentally measured transmission coefficients $|T|$ (black solid line) by \citet{zhengwu2025} with our numerical predictions from Model 1 (red solid line) and Model 2 (blue dashed line) for a 15~Hz capillary–gravity wave scattered by a 3.18~mm-wide surface-piercing plate barrier. Initially, the barrier is positioned so that the contact angle sets $\theta = 180^\circ$, and raising the barrier reduces $\theta$ from hydrophobic ($>90^\circ$) to $90^\circ$ (flat meniscus) and then to $0^\circ$ (positively curved meniscus). Across this range, $|T|$ increases steadily in the experiment, and both Model 1 and Model 2 capture this trend accurately, showing how moderate meniscus curvature enhances transmission beyond what a flat interface would predict.

Notably, the inset of Fig.~\ref{fig:exp_vs_sim}, which shows the absolute and relative differences between the two models, follows from the analysis in Sec.~\ref{sec:Bridge}. This confirms that the observed discrepancy indeed stems from the distinct governing equations in Model 2 compared to Model 1. Moreover, as shown in the inset of Fig.~\ref{fig:exp_vs_sim}, the difference between the two models’ results is extremely small. The maximum difference between the models is 0.02, and the maximum relative difference is 5\%. When the meniscus is sufficiently flat \(\theta \approx 90^\circ \), the relative difference approaches to zero, and Model 1 coincides with Model 2, as predicted in Sec.~\ref{sec:Bridge}.

However, once $\theta$ becomes negative, the meniscus is overturned (curved back). Because Model 1 relies on a single‐valued representation $y = \eta(x)$, it cannot handle these overturned (multi‐valued) interfaces and fails to capture the abrupt decline in $|T|$. By contrast, Model 2, which admits arbitrary surface geometries, reproduces the rapid decrease in $|T|$ beyond $\theta = 0^\circ$. Even after the meniscus would have broken in the laboratory, Model 2 can formally continue to $\theta = -90^\circ$, predicting $|T| \approx 0$ near $\theta = -65^\circ$ and a rebound as $\theta$ decreases further to $-90^\circ$. Although such states lie outside the experimentally realizable range, they demonstrate Model 2’s robustness for strongly curved or overturned surfaces.

\subsection{Field Visualizations and Interpretations}

In the experiment, only the transmitted wave elevation is recorded, whereas the simulation provides a more comprehensive view of the entire flow field. Figure~\ref{fig:fields} highlights how the meniscus geometry and contact angle $\theta$ affect both the wave profile and the underlying flow in Model~2. The unperturbed meniscus $\eta_0$ is indicated by the dashed black curve, whereas the perturbed (wave‐bearing) surface $\eta$ is shown by the solid black curve. The barrier is drawn in the center, and the color scale denotes the normalized velocity potential field $\phi/\phi_A$, where red/blue represent positive/negative values. From Fig.\ref{fig:fields}, we note the following key observations, aligned with the interpretations in \citet{zhengwu2025}:

\begin{itemize}
\item \(\theta = 180^\circ\): The disturbance of the downstream side free surface is slight, showing the barrier size effect in blocking the transmission of capillary-gravity wave is significant when the barrier is below the water level.
\item Transition from \(\theta = 180^\circ\) to \(\theta = 90^\circ\): As the meniscus evolves from a negative shape to a nearly flat interface, the wave amplitude on the upstream side grows, and the overall transmission rises. Relieving the meniscus deformation increases the effective coupling between the upstream and downstream flows.
\item Transition from \(\theta = 90^\circ\) to \(\theta = 0^\circ\): Transmission continues to increase as a positively curved meniscus develops. The forming of the water column beneath the barrier enhances the coupling of the flow from the incident side to the transmitted side, leading to the increase of the transmission.
\item Transition from \(\theta = 0^\circ\) to \(\theta = -65^\circ\): Once $\theta$ goes negative, the meniscus overturns, creating a pronounced dip that can significantly alter the flow beneath the free surface. The meniscus becomes so inclined to constrain the surface motion and thus enhance the reflection. At the specific angle $\theta=-65^\circ$, as shown in Fig.~\ref{fig:exp_vs_sim}, the simulation predicts zero wave transmission, a result also evident in the color map of Fig.~\ref{fig:fields}: the velocity potential to the right of the barrier (downstream) is nearly white (indicating negligible wave energy), indicating complete reflection. 
\item Transition from \(\theta = -65^\circ\) to \(\theta = -90^\circ\): Decreasing \(\theta\) further curves the overturned meniscus and thinner the distance between the two meniscus, allowing some downstream wave energy to reemerge and modestly boosting the transmission, which is an effect not observed experimentally. By \(\theta=-90^\circ\), a deep cusp has formed, yet there is a modest return of wave activity downstream. 
\end{itemize}

Overall, these results underscore the critical impact of meniscus curvature in small‐scale wave‐structure interactions and demonstrate the advantages of modeling the surface displacement in the direction normal to the unperturbed interface (Model 2) when overturning or extreme curvature arises. Model 1 remains accurate for gentle menisci but breaks down once the interface ceases to be single‐valued in $x$.

\section{Summary and Conclusion}
\label{sec:Summary}

In this paper, we investigated how meniscus curvature near a surface-piercing barrier affects the scattering of capillary–gravity waves—particularly in regimes where classical flat-surface assumptions fail to capture experimentally observed behavior. Recent laboratory studies demonstrated a notable surge in transmitted wave energy when the barrier intersects the meniscus at its crest, followed by a swift drop as the barrier is raised further and the meniscus becomes overturned or breaks. Motivated by these findings, we developed two theoretical models that incorporate meniscus geometry into the modeling of linearized capillary–gravity wave propagation and scattering.

The first model extends the standard approach of measuring wave elevations in the vertical direction while retaining the static meniscus profile \(y = \eta_0(x)\). We derived linearized surface boundary conditions for small-amplitude wave perturbations \(\eta_1\) traveling on the meniscus  \(\eta_0(x)\), which reduces to the well-known capillary–gravity wave equations if \(\eta_0=0\) (a flat interface). This formulation is mathematically straightforward and accurate when the meniscus remains single-valued in \(x\). However, because vertical displacements cannot accommodate large slopes or overturning, the model is limited in scenarios where the interface becomes multi-valued.

To overcome this geometric restriction, we introduced a second model that measures wave perturbations perpendicular (normal) to the meniscus. By parameterizing the free surface in terms of arc length, this model naturally incorporates arbitrary meniscus shapes, including those that loop back or exhibit vertical tangents. Model 2 reduces to Model 1 when the free surface is single-valued and the meniscus slope is small. Outside this small-slope regime, Model 2 is capable to capture the influence of steep or overturned menisci on wave propagation.

Analytical comparisons reveal that the general (fully nonlinear) boundary conditions in Model 2 and Model 1 are formally equivalent for single-valued surfaces, but their linearized versions differ by a small extra term related to the meniscus curvature. This term remains negligible when the meniscus is nearly flat. Therefore, Model 1 suffices to predict transmission trends at small-slope meniscus. Once the meniscus becomes appreciably curved or overturned, however, Model 2 is indispensable.

To validate these formulations, we simulated a 15 Hz capillary–gravity wave encountering a 3.18 mm-wide barrier—a setup previously studied in experiments. We compared the numerically predicted transmission coefficient \(\lvert T\rvert\) against experimental measurements across a range of meniscus contact angles \(\theta\). Model 1 reproduced the initial rise in \(\lvert T\rvert\) as \(\theta\) decreased from \(113.6^\circ\) (a negative meniscus) down to \(0^\circ\) (a positive meniscus), matching experimental observations for gentle menisci. However, it could not capture the sudden drop in \(\lvert T\rvert\) that occurs once \(\theta\) becomes negative and the meniscus overturns. By contrast, Model 2 closely tracked the experimental transmission behavior in both the gentle-slope and overturned-slope regimes, showing a precipitous decline in wave transmission until meniscus breakup and even beyond the experimentally observed limits.

These results underscore the critical role of the meniscus in small-scale wave–structure interactions. By accounting for contact line effects and surface curvature, the proposed models provide a more comprehensive framework than classic flat-surface analyses. Model 1 offers computational simplicity and clear parallels to standard capillary–gravity theory when meniscus slopes are mild. Model 2, while more complex, captures the full geometry of the interface—including overturned surfaces—and therefore remains valid in a wider range of contact angles. While a steep meniscus bends the surface to suppress the transmission, Model 2 even predicted that at certain slope the water column underneath the barrier becomes so thin to enhance the coupling from one side to the other, which was not observed in experiments due to the breaking of the water column. Taken together, these two models constitute a robust theoretical foundation for predicting capillary-gravity wave propagation and scattering in systems where meniscus presents to alter fluid behavior. 

While the models successfully replicate the measured transmission trends for a given wave frequency and fixed barrier width, the variation of transmission with frequency and barrier width requires further investigation to deepen the understanding of meniscus effects. Although the present work accounts for the meniscus geometry, it assumes a \emph{pinned} contact line, neglecting the possibility that the contact line itself might move. Extending the models to incorporate a \emph{dynamic} contact line, following approaches similar to \citet{guoqin2025}, would allow for more realistic simulations of wave scattering under changing wetting conditions. Addressing these issues will help clarify the fundamental physics of wave–structure interactions in small‐scale, surface‐tension‐dominated flows and could guide the design of engineered systems that exploit or mitigate capillary–gravity effects.

\backsection[Supplementary data]{\label{SupMat}Supplementary material and movies are available at \\https://doi.org/10.1017/jfm.2019...}

\backsection[Funding]{Please provide details of the sources of financial support for all authors, including grant numbers. Where no specific funding has been provided for research, please provide the following statement: "This research received no specific grant from any funding agency, commercial or not-for-profit sectors." }

\backsection[Declaration of interests]{The authors report no conflict of interest.}

\backsection[Author ORCIDs]{G. Liu, https://orcid.org/0009-0002-1452-2267; Z. Wang, https://orcid.org/0009-0009-3252-4646; L. Zhang, https://orcid.org/0000-0003-3898-6533}

\appendix

\section{Justification for reusing the arc length}\label{app:arclength}

In this appendix, we justify the feasibility of reusing the arc length parameter in \eqref{eq:pert_M2}:
\begin{equation}
    \bs(s, t) = \bs_0(s) + \eta_1(s, t) \nhat_0(s), 
\end{equation} 
that when the surface perturbation $\eta_1$ is sufficiently small, the arc length of the perturbed surface $\bs$ differs from that of the unperturbed surface $\bs_0$ by a term of higher (second) order in $\eta_1$.

Let $s$ be the arc length of the unperturbed surface $\bs_0$. This means:
\begin{equation}
    \left\Vert \bs_0'(s) \right\Vert = \sqrt{\soxp^2 + \soyp^2} = 1.
\end{equation}
In 2D, $\bs_0'(s)$ is simply the unit tangent, which is perpendicular to $\nhat_0(s)$.

Differentiate $\bs(s, t)$ with respect to $s$:
\begin{align}
    \bs'(s, t) &= \bs_0'(s) + \eta_1'(s, t) \nhat_0(s) + \eta_1(s, t) \nhat_0'(s), \nonumber\\
    &= \bs_0' + \eta_1' \nhat_0 - \eta_1 \kappa_0 \bs_0', \nonumber\\
    &= (1 - \eta_1 \kappa_0) \bs_0' + \eta_1' \nhat_0.
\end{align}

Consider the norm of $\bs'$:
\begin{align}
    \left\Vert \bs'(s, t) \right\Vert &= \left\Vert (1 - \eta_1 \kappa_0) \bs_0' + \eta_1' \nhat_0 \right\Vert, \nonumber\\
    &= \sqrt{(1 - \eta_1 \kappa_0)^2 \left\Vert \bs_0' \right\Vert^2 + \eta_1'^2 \left\Vert \nhat_0 \right\Vert^2 + 2 (1 - \eta_1 \kappa_0) \eta_1' \bs_0' \bcdot \nhat_0}. 
\end{align}
Notice that $\left\Vert \bs_0' \right\Vert = 1$, $\left\Vert \nhat_0 \right\Vert = 1$, and $\bs_0' \bcdot \nhat_0 = 0$, hence the above expression simplifies to:
\begin{equation}
    \left\Vert \bs'(s, t) \right\Vert = \sqrt{1 - 2 \kappa_0 \eta_1 + \kappa_0^2\eta_1^2 + \eta_1'^2}.
\end{equation}
Expand this norm to first order in $\eta_1$:
\begin{equation}
    \left\Vert \bs'(s, t) \right\Vert = 1 - \kappa_0 \eta_1 + \mathcal{O}(\eta_1^2).
\end{equation}
Thus, if $\kappa_0 \eta_1 \ll 1$, $\left\Vert \bs' \right\Vert = \left\Vert \bs_0' \right\Vert = 1$. The difference between the arc length of the perturbed surface $\bs$ and the unperturbed surface $\bs_0$ is $\mathcal{O}(\kappa_0 \eta_1)$. For meniscus surface influenced by surface tension and gravity, the surface curvature $\kappa_0$ is of the order of $1/a$, where $a = \sqrt{\sigma/\rho g}$ is the capillary length. 

In summary, if $\eta_1$ satisfies the smallness relation:
\begin{equation}
    \eta_1 \ll a,
\end{equation}
we can use the arc length $s$ for the unperturbed surface $\bs_0$ as the parameter for the perturbed surface $\bs$.

\section{Meniscus profile parameterized by arc length}\label{app:Meniscus_M2}

In this appendix, we derive an explicit expression for the meniscus profile $(\sx(s), \sy(s))$ near a vertical wall with contact angle $\theta$, where the free surface is parameterized by arc length $s$. For the right-going meniscus (Fig.~\ref{fig:meniscus}(b)), the functions $\sx(s)$ and $\sy(s)$ satisfy \eqref{eq:sp_relation} and \eqref{eq:dynamic_app2_M2} which forms the following boundary value problems: 
\begin{subequations}\label{eq:app1_eq1}
    \begin{empheq}[left={\empheqlbrace}]{align}
        &\sxp^2 + \syp^2 = 1, \label{eq:app1_eq1a}\\
        &\sxp \sypp - \syp \sxpp = a^{-2} \sy, \label{eq:app1_eq1b}\\
        &\sx(0) = 0,\ \syp(0) = -\cos\theta,\ \sy(+\infty) = 0,\ \syp(+\infty) = 0, \label{eq:app1_eq1c}
    \end{empheq}
\end{subequations}
where $a = \sqrt{\sigma/\rho g}$ is the capillary length, $\sxp = \ps \sx$, and $\syp = \ps \sy$.

To decouple \eqref{eq:app1_eq1a} and \eqref{eq:app1_eq1b}, note from \eqref{eq:app1_eq1a} that:
\begin{equation}
    \sxp = \sqrt{1 - \syp^2}. \label{eq:app1_eq2}
\end{equation}
The positive square root is chosen because $\sx$ increases with $s$. Substitute \eqref{eq:app1_eq2} into \eqref{eq:app1_eq1b} yields the decoupled second-order ODE for $\sy(s)$:
\begin{equation}
    \sypp = a^{-2} \sy \sqrt{1 - \syp^2}, \label{eq:app1_eq3}
\end{equation}
which can be integrated once to give:
\begin{equation}
    -\sqrt{1 - \syp^2} = \frac{\sy^2}{2 a^2} + C, \label{eq:app1_eq4}
\end{equation}
where $C$ is an integration constant. The boundary conditions at $s \to +\infty$ ($\sy \to 0$ and $\syp \to 0$) imply $C = -1$. \eqref{eq:app1_eq4} then reduce to a first-order ODE:
\begin{equation}
    \syp = - \frac{\sy \sqrt{4 a^2 - \sy^2}}{2 a^2}, \label{eq:app1_eq5}
\end{equation}
where the negative sign ensures that $\sy$ and $\syp$ have opposite signs. 

Next, set:
\begin{equation}
    \sy = 2 a \sin \xi. \label{eq:app1_eq6}
\end{equation}
Then \eqref{eq:app1_eq5} becomes:
\begin{equation}
    \xi' = - a^{-1} \sin\xi, \label{eq:app1_eq7}
\end{equation}
with the general solution:
\begin{equation}
    \tan(\xi/2) = \A \exp (-s/a), \label{eq:app1_eq8}
\end{equation}
where $\A$ is a constant to be determined. Substituting \eqref{eq:app1_eq8} back into \eqref{eq:app1_eq6} yields the expression for $\sy(s)$:
\begin{equation}
    \boxed{
    \sy(s) = \frac{4 a \A \ue^{-s/a}}{1 + \A^2 \ue^{-2 s/a}}. \label{eq:app1_y}
    }
\end{equation}
Applying the boundary condition $\syp(0) = -\cos \theta$ at $s = 0$ gives $\A = \tan(\tfrac{\pi}{8} - \tfrac{\theta}{4})$. The curvature $\kappa(s)$ follows as:
\begin{equation}
    \boxed{
    \kappa(s) = \sxp \sypp - \syp \sxpp = a^{-2} \sy = \frac{4 a^{-1} \A \ue^{-s/a}}{1 + \A^2 \ue^{-2 s/a}}. \label{eq:app1_kappa}
    }
\end{equation}
Finally, substituting \eqref{eq:app1_y} into \eqref{eq:app1_eq2} and integrate $\sxp$ from $0$ to $s$ provides the expression for $\sx(s)$ of a right-going meniscus:
\begin{equation}
    \sx(s) = s + 4 a \left[ \frac{1}{1 + \A^2} - \frac{1}{1 + \A^2 \ue^{-2 s/a}}\right]. \label{eq:app1_x_temp}
\end{equation}
Given that the left-going meniscus is a mirror image of the right-going meniscus with respect to the $y$-axis, $\sy(s)$ and $\kappa(s)$ for the left-going meniscus will still be \eqref{eq:app1_y} and \eqref{eq:app1_kappa}, while $\sx(s)$ for the left-going meniscus will have different sign as \eqref{eq:app1_x}. Therefore, the general expression for $\sx(s)$ will be:
\begin{equation}
    \boxed{
    \sx(s) = \pm \left\{ s + 4 a \left[ \frac{1}{1 + \A^2} - \frac{1}{1 + \A^2 \ue^{-2 s/a}}\right] \right\}. \label{eq:app1_x}
    }
\end{equation}

In summary, Eqs.~\eqref{eq:app1_x}, \eqref{eq:app1_y}, and \eqref{eq:app1_kappa} fully specify the meniscus shape $(\sx(s),\sy(s))$ along a vertical wall with contact angle $\theta$, parameterized by arc length $s$.

\end{document}